\documentclass[%
superscriptaddress,
preprint,
amsmath,amssymb,
]{revtex4-1}

\usepackage{amsmath}
\usepackage{amssymb}%
\usepackage{natbib}
\usepackage{graphicx}
\usepackage{dcolumn}
\usepackage{bm}
\usepackage{hyperref}
\usepackage[caption=false]{subfig}
\usepackage{ulem,color}
\usepackage[outdir=./]{epstopdf}

\newcommand{\lt}{\left}
\newcommand{\rt}{\right}

\newcommand{\mrm}{\mathrm}
\newcommand{\mbf}{\mathbf}
\newcommand{\mcl}{\mathcal}
\newcommand{\mbb}{\mathbb}

\newcommand{\der}{\mrm{d}}

\newcommand{\im}{\mrm{i}}

\begin{document}
	\title{Majorana Zero-Energy Mode and Fractal Structure in Fibonacci--Kitaev Chain}
	\author{Rasoul Ghadimi}
	\affiliation{Department of Physics, Sharif University of Technology, Tehran 11155-9161, Iran}
	\affiliation{Department of Applied Physics, Tokyo University of Science, Tokyo 125-8585, Japan}
	\author{Takanori Sugimoto}
	\affiliation{Department of Applied Physics, Tokyo University of Science, Tokyo 125-8585, Japan}	
	\author{Takami Tohyama}
	\affiliation{Department of Applied Physics, Tokyo University of Science, Tokyo 125-8585, Japan}
	\date{\today}
\begin{abstract}
	We theoretically study a Kitaev chain with a quasiperiodic potential, where the quasiperiodicity is introduced by a Fibonacci sequence.
	Based on an analysis of the Majorana zero-energy mode, we find  the critical $p$-wave superconducting pairing potential separating a topological phase and a non-topological phase.
	The topological phase diagram with respect to Fibonacci potentials follow a self-similar fractal structure characterized by the box-counting dimension, which is an example of the interplay of fractal and topology like the Hofstadter's butterfly in quantum Hall insulators.
\end{abstract}
\maketitle
\section{Introduction}
Since the discovery of the Berezinskii--Kosterlitz--Thouless transition~\cite{Berezinskii71,Kosterlitz73}, topology has been a potential probe to clarify a universal (but sometimes hidden) character of quantum phenomena in condensed matters.
Topological phenomena have been extensively studied for various systems, not only theoretically but also experimentally, such as the quantum Hall effect in two-dimensional electron systems~\cite{Klitzing80, Thouless82} and the Berry phase in the Haldane state~\cite{Haldane83, Affleck87, Hirano08}.
Recent topological researches are motivated by potential application to quantum computation by regarding a topological object as a qubit.
The qubit can be tolerant to external perturbations in contrast to a usual qubit that is fragile with respect to some perturbations like impurities and external fields.
As one of the characteristics of such a topological object, a zero-energy mode of Majorana fermions localized on different edges of the Kitaev chain has intensively studied in the last decade~\cite{Kitaev01,Wilczek09}.

The zero-energy mode of Majorana fermions in the Kitaev chain appears if there is an infinitesimally small $p$-wave superconducting pairing potential, provided that the uniform chemical potential is within the bandwidth. When the random potential is introduced in the chain, there is a finite critical value of the pairing potential below which the system is non-topological without the Majorana zero-energy mode~\cite{DeGottardi13}. 

In addition to random potentials, other types of aperiodic potential are possible in the Kitaev chain.
Since one-dimensional quasiperiodic systems are now experimentally accessible by ultracold atoms~\cite{Modugno10,Kraus16}, the Kitaev chain with quasiperiodicity is an interesting system.
The quasiperiodicity can be introduced by the Harper model~\cite{Harper55}, corresponding to the Hofstadter model~\cite{Azbel64,Hofstadter76} in two-dimensional square lattice with a magnetic flux on each plaquette.
The Hofstadter model induces a fractal structure represented by the Hofstadter's butterfly in the electron spectrum.
The Kitaev chain with the Harper potential has been studied and the Hofstadter's butterfly has beautifully observed in the distribution of the inverse of the localization length~\cite{DeGottardi13}.

Another possible case for quasiperiodicity is the potential with Fibonacci sequence. 
The Fibonacci potential is a superposition of an infinite number of the Harper potentials~\cite{Naumis08}, and its topological phase is smoothly connected to that of the Harper model~\cite{Kraus12}.
However, the Fibonacci potential is different from a superposition of a finite number of the Harper potentials in terms of real-space self-similarity.
In addition, since it has been reported that such a real-space self-similarity results in a self-similar structure (a Cantor set) in the energy spectrum~\cite{Kohmoto83,Kohmoto87}, the Fibonacci potential introduced into the Kitaev chain also has a possibility to produce new self-similar fractal structures with relation to the Majorana zero-energy mode.
A pioneering work has indicated the presence of fractal structures in the wave functions in such a model~\cite{Satija13}. Further detailed studies to characterize the fractal structures are desired to fully understand the interplay of fractal and topology in the Fibonacci--Kitaev chain. 

In this work, we perform a detailed study on the Majorana zero-energy mode in the Kitaev chain with the Fibonacci potential.
Based on an analysis of the Majorana zero-energy mode, we find a phase transition between a topological phase and a non-topological phase in terms of localization of the fermions.  
Examining the critical values of $p$-wave pairing potential above which topological phase emerges, we find a self-similar fractal structure in the topological phase diagram with respect to the Fibonacci potentials. We estimate the box-counting dimension $D\approx 1.7$, which characterizes the self-similar fractal structure~\cite{Mandelbrot77}. This will provide useful information available for future experimental confirmation of the interplay of fractal and topology in condensed matter physics.


\begin{figure}[t]
	\centering
	\includegraphics[width=0.48\textwidth]{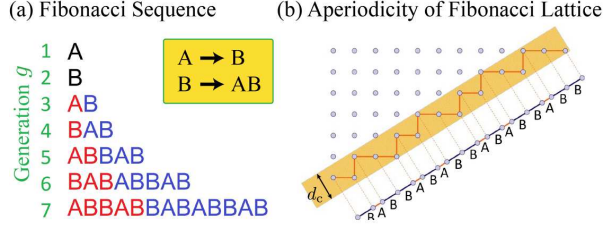}
	\caption{(Color online)(a) Fibonacci sequence as a function of generation $g$. By applying an inflation rule $B\rightarrow AB$, $A\rightarrow B$ (represented in yellow square) to the $g$th generation of Fibonacci chain, we obtain the $(g+1)$th generation. The red and blue letters denote the parts descended from A and B sites of the 3rd generation, respectively. The system size of the $g$th generation $N_g$ obeys the Fibonacci recurrence formula $N_g=N_{g-1}+N_{g-2}$.
		(b) Quasiperiodicity of the Fibonacci chain obtained by cutting and projecting approach from the square lattice. This approach gives the same structure as the Fibonacci sequence.
		The cutting depth is denoted by $d_{\mrm{c}}$ and the projection is done perpendicular to the chain.}
	\label{fig:fibo}
\end{figure}

\section{\label{model} Model and Approach}

We consider the Kitaev chain with a quasiperiodic on-site chemical potential $\mu_i$,
\begin{equation}
\mathcal{H}=\sum_{i=1}^{N-1} (-c^\dagger_ic_{i+1}+{\it\Delta} c_ic_{i+1}+\mrm{H.c.}) +\sum_{i=1}^N\mu_i c^{\dagger}_ic_i
\label{eq:ham}	
\end{equation}
where $c^\dagger_i$ ($c_i$) is a creation (an annihilation) operator of spinless fermion on site {\it i}, and ${\it\Delta}$ is a {\it p}-wave superconducting pairing potential in the $N$ sites system.  
The Kitaev chain ($\mu_i=0, {\it\Delta}\neq0$) is a typical model exhibiting a topological phenomenon, i.e., Majorana zero-energy mode.
We note that the topological phase is extended to $|\mu_i|<2$ with ${\it\Delta}\neq 0$ if the on-site potentials are uniform~\cite{Hassler12}.
The quasiperiodicity in $\mu_i$ is given by assigning each $\mu_i$ by either $\mu_{\mrm{A}}$ or $\mu_{\mrm{B}}$, where the order of A and B is introduced by Fibonacci sequence.
In the Fibonacci sequence, the first generation of Fibonacci string is given by only one character A [Fig.~\ref{fig:fibo}(a)].
The following generations are obtained by using a substitution rule A$\to$B and B$\to$AB step by step, and thus the system size $N_g$ as a function of the generation $g$ is given by the Fibonacci series $N_g=N_{g-1}+N_{g-2}$.
By definition, a high generation of Fibonacci string has a self-similar fractal structure.
The quasiperiodicity of this system is confirmed if we consider a skewed two-dimensional (2D) square lattice [Fig.~\ref{fig:fibo}(b)].
We can also obtain the Fibonacci string by using a finite cutoff of 2D square lattice and a projection onto one dimension~\cite{Kraus16}.

To clarify a topological transition and determine its boundary, we examine a Majorana zero-energy mode by using a conventional technique.
The Fibonacci--Kitaev chain (\ref{eq:ham}) can be mapped to Majorana-fermion Hamiltonian by introducing two Majorana fermions $a_j$ and $b_j$ with $c_j=(a_j+\im b_j)/2$~\cite{Kitaev01}:
In the topological phase of this system, we can find a zero-energy mode composed by two Majorana fermions, $Q_{\mrm{a}}=\sum_i\alpha_i a_i$ and $Q_{\mrm{b}}=\sum_i\beta_i b_i$, where $\alpha_i$ ($\beta_i$) is the amplitude of the superposition $a_i$ ($b_i$). 
The conditions for the presence of zero-energy mode are as follows~\cite{DeGottardi13,DeGottardi131,DeGottardi11}: (I) $Q_{\mrm{a}}$ and $Q_{\mrm{b}}$ commute with Hamiltonian, and (II) $Q_{\mrm{a}}$ and $Q_{\mrm{b}}$ are normalizable even if the system size is infinite. 
The condition (I) corresponds to the following equation: 
\begin{equation}
\left(\begin{matrix}
\alpha_{i+1}\\\alpha_i
\end{matrix}\right)=\mbf{A}_i
\left(\begin{matrix}
\alpha_{i}\\\alpha_{i-1}
\end{matrix}\right)
\text{ with }
\mbf{A}_i=\left(
\begin{matrix}
\frac{\mu_i}{1+{\it\Delta}} & \frac{{\it\Delta}-1}{1+{\it\Delta}} \\ 1 & 0
\end{matrix}
\right),
\label{eq:tm}
\end{equation}
where $\mbf{A}_i$ is the transfer matrix of site \textit{i} (see Appendix for deriving this equation).
The normalization condition (II) can be checked by the number $v\equiv (-1)^{n_{\mrm{f}}-1}$, where $n_{\mrm{f}}$ is the number of eigenstates of $\mbf{\Lambda}^g\equiv\prod_{i=1}^{N_g}\mbf{A}_i$ whose eigenvalues are smaller than 1.
If $v=-1$ ($+1$), the normalization condition (II) is (not) satisfied~\cite{note}.
In fact, the zero-energy mode is localized near the edge of the lattice, when both of the two eigenvalue of $\mbf{\Lambda}^g$ are smaller or larger than 1.
Therefore, $v$ is a topological invariant, and $v=-1$ ($+1$) means that system is topological (non-topological). 
If the absolute value of the two eigenvalues of $\mbf{\Lambda}^g$ is one, the system stands on its topological phase boundary, where an energy gap vanishes~\cite{DeGottardi13,DeGottardi11}.

To determine the topological invariant $v$, we define $\lambda_1$ and $\lambda_2$ as the smaller and larger eigenvalues of $\mbf{\Lambda}^g$, respectively. 
In the following, we assume ${\it\Delta}\geq0$, which does not lose the generality.
Since $\mrm{det}[\mbf{\Lambda}^g]=\lt(\frac{1-{\it\Delta}}{1+{\it\Delta}}\rt)^{N_g} \to 0$ as $N_g\to\infty$ with ${\it\Delta}>0$, $|\lambda_1|$ exponentially decreases so that $v=\mrm{sgn}(\gamma^{g})$ with $\gamma^{g}(\{\mu_i\},{\it\Delta})\equiv \frac{1}{N_g} \mrm{ln}|\lambda_2(\{\mu_i\},{\it\Delta})|$, where $\{\mu_i\}$ is a set of on-site potential.
We call $\gamma^g$ the Lyapunov exponent according to Refs.~[10,22]. 
The coefficient $\frac{1}{N_g}$ in $\gamma^g$  means that $\gamma^g$ has a non-zero value if $|\lambda_2|$ is of the order of $e^{\eta N_g}$. If $\eta<0$ and $|\lambda_2|\ll 1$, $\gamma^g<0$, which implies that the system has a well-defined localized mode near the edges.
Vanishing the Lyapunov exponent in the thermodynamical limit, $\lim_{g\to\infty} \gamma^g=0$, is equivalent to the condition of topological phase boundary, and a negative (positive) value means that the phase is topological (non-topological) in the thermodynamical limit. 	
Assuming $0<{\it\Delta}<1$, $\gamma^g$ satisfies the following equation (see Appendix for the derivation),
\begin{equation}
\gamma^g(\{\mu_i\},{\it\Delta})=\gamma^g\lt(\lt\{\frac{\mu_i}{\sqrt{1-{\it\Delta}^2}}\rt\},0\rt)-\frac{1}{2}\mrm{ln}\lt(\frac{1+{\it\Delta}}{1-{\it\Delta}}\rt).
\label{eq:lyup}
\end{equation}
The Lyapunov exponent $\gamma^g(\{\mu_i\},{\it\Delta})$ is, thus, given by $\gamma_0^g\lt(\{\mu_i\}\rt)\equiv\gamma^g(\{\mu_i\},0)$ with (i) a rescale of $\mu_i$ to $\frac{\mu_i}{\sqrt{1-{\it\Delta}^2}}$ and (ii) a shift of $\gamma_0^g$ by $\frac{1}{2}\mrm{ln}(\frac{1+{\it\Delta}}{1-{\it\Delta}})$.

\section{\label{Numerical}Numerical Results}
\begin{figure}
	\centering
	\includegraphics[width=0.5\textwidth]{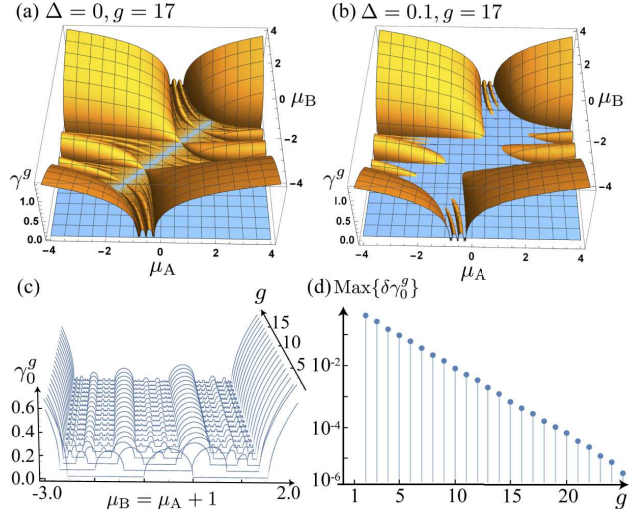}
	\caption{(Color online)Lyapunov exponent $\gamma^{g}$. The region of $\gamma^g<0$ ($\gamma^g>0$) is topological (non-topological).  
		(a) Three-dimensional plot of $\gamma_0^g$ for $g=17$ without superconducting pairing potential (${\it\Delta}=0$) as a function of potentials, $\mu_A$ and $\mu_B$. The blue surface denotes the $\gamma^g=0$ plane. The diagonal blue line denotes the region with uniform chemical potential. 
		(b) The same as (a) but for ${\it\Delta}=0.1$. 
		(c) $\gamma_0^g$ as a function of $\mu_A=\mu_B-1$ for various generation $g$. 
		(d) Semi-log plot of the maximal difference of $\gamma_0^g$ as compared with that of the previous generation, i.e., $\delta \gamma_0^g=|\gamma_0^g-\gamma_0^{g-1}|$.
	}
	\label{fig:lyap}
\end{figure}

Figures~\ref{fig:lyap}(a) and \ref{fig:lyap}(b) show $\gamma^{g=17}$ with ${\it\Delta}=0$ and $0.1$, respectively, as a function of $\mu_A$ and $\mu_B$. 
We find an intricate structure like a ravine, where a central region in Fig.~\ref{fig:lyap}(a) (light blue region) sinks below the zero surface in Fig.~\ref{fig:lyap}(b) due to the effect (ii).
Though it is hard to see the effect (i) in Figs.~\ref{fig:lyap}(a) and \ref{fig:lyap}(b), we can confirm the effect (i) with a large ${\it\Delta}$ close to 1 (not shown).
We investigate the generation $g$ dependence of the Lyapunov exponent $\gamma_0^{g}$ with fixed potentials $\mu_{\mrm{B}}=\mu_{\mrm{A}}+1$.
As we can see in Fig.~\ref{fig:lyap}(c), $\gamma_0^g$ converges into a finite value in the thermodynamical limit $g\to\infty$.
Actually, the maximal value of the difference $\delta \gamma_0^g=|\gamma_0^g-\gamma_0^{g-1}|$ exponentially decreases with increasing $g$ as shown in Fig.~\ref{fig:lyap}(d), indicating the convergence of the Lyapunov exponent in the thermodynamical limit~\cite{note2}.

\begin{figure}[t]
	\centering
	\includegraphics[width=0.4\textwidth]{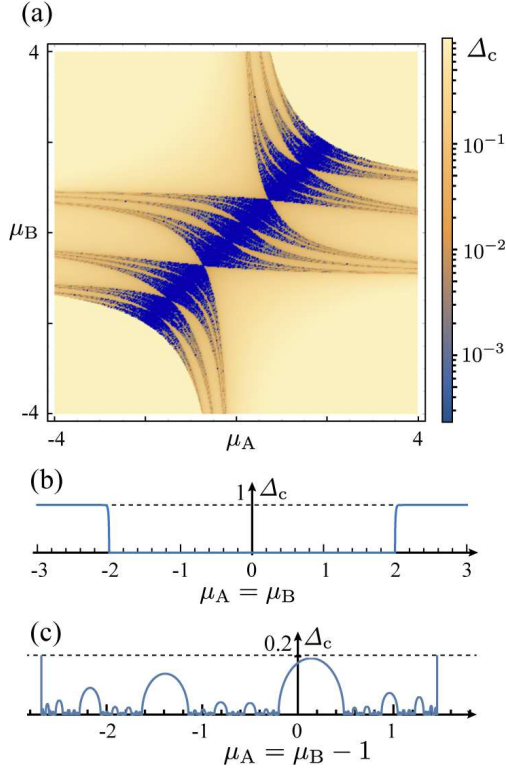}
	\caption{(Color online)(a) Critical $p$-wave pairing potential ${\it\Delta}_{\mrm{c}}$ of the Fibonacci--Kitaev model for the $g=17$ generation of the Fibonacci sequence. Below ${\it\Delta}_{\mrm{c}}$, the system is a non-topological phase.
		We put ${\it\Delta}_{\mrm{c}}=1$ in the region centered at $\mu_A=\mu_B=\pm 4$ and surrounded by the outermost arc, since there is no self-consistent solution of ${\it\Delta}_{\mrm{c}}$. This indicates that ${\it\Delta}_{\mrm{c}}$ in this region is out of the assumed range ${\it\Delta}<1$.
		(b) ${\it\Delta}_{\mrm{c}}$ along the $\mu_A=\mu_B$ line (uniform on-site potential) in (a).  
		(c) ${\it\Delta}_{\mrm{c}}$ along the $\mu_A=\mu_B-1$ line in (a). ${\it\Delta}_{\mrm{c}}$ shows a fractal structure.}
	\label{fig:cp}
\end{figure}

We switch to the effects of the superconducting pairing potential ${\it\Delta}$.
Since the phase boundary between the topological and non-topological phases is given by $\gamma^g=0$, a critical paring potential ${\it\Delta}_{\mrm{c}}$ should satisfy a self-consistent equation: ${\it\Delta}_{\mrm{c}}=\tanh\gamma_0^g(\{\frac{\mu_i}{\sqrt{1-{\it\Delta}_{\mrm{c}}^2}}\})$ [see Eq.~(\ref{eq:lyup})].
Figure~\ref{fig:cp}(a) shows ${\it\Delta}_{\mrm{c}}$ as a function of $\mu_{\mrm{A}}$ and $\mu_{\mrm{B}}$ in the $g=17$ generation.
Along the $\mu_{\mrm{A}}=\mu_{\mrm{B}}$ line, where the on-site potential is uniform, ${\it\Delta}_{\mrm{c}}=0$ for $|\mu_{\mrm{A}}|<2$, as depicted in Fig.~\ref{fig:cp}(b), i.e., an infinitesimal pairing potential ${\it\Delta}$ ($>0$) induces a topological phase as is well-known for the Kitaev chain. 
The region of $0\le {\it\Delta}_{\mrm{c}}<1$ is surrounded by the two outermost arcs centered at $\mu_A=\mu_B=\pm4$ as seen in Fig.~\ref{fig:cp}(a).
Note that ${\it\Delta}_{\mrm{c}}\ge 1$ out of the region, where Eq.~(\ref{eq:lyup}) cannot be applied. 
In the region of $0\le {\it\Delta_{\mrm{c}}}<1$, ${\it\Delta}_{\mrm{c}}$ exhibits an intricate structure. For example, ${\it\Delta}_{\mrm{c}}(\{\mu_i\})$ along $\mu_{\mrm{A}}=\mu_{\mrm{B}}-1$ is shown in Fig.~\ref{fig:cp}(c). The intricate structure also appears in the energy spectrum and in the wave function distribution as is Ref.~[19] (not shown).
Therefore, the structure is expected to exhibit a fractal nature, as confirmed below.

\section{Fractal Structure}
In the following, we discuss the intricate structure of ${\it\Delta}_{\mrm{c}}$ and characterize the fractal by the box-counting dimension~\cite{Mandelbrot77}.
For ${\it\Delta}=0$, the eigenvalues of $\mbf{\Lambda}^g$, i.e., $\lambda_1$ and $\lambda_2$, satisfy the relation $\lambda_1=\lambda_2^{-1}$ because $\mrm{det}[\mbf{\Lambda}^g]=1$. We thus obtain $\lambda_2=\left(|\mrm{Tr}[\mbf{\Lambda}^g]|+\sqrt{(\mrm{Tr}[\mbf{\Lambda}^g])^2-4}\right)/2$. 
Consequently, $\gamma_0^g$ is given by
\begin{equation}
\gamma_0^g=\left\{\begin{matrix}
N_g^{-1}\cosh^{-1} (|\frac{1}{2}\mrm{Tr}[\mbf{\Lambda}^g]|) & \lt(|\mrm{Tr}[\mbf{\Lambda}^g]|> 2\rt)\\
0 & \lt(|\mrm{Tr}[\mbf{\Lambda}^g]|\le 2\rt)
\end{matrix}\right. .
\label{eq:g0g}
\end{equation}
Since the first and second generations of the Fibonacci sequence have just one site, A and B, respectively, $\mrm{Tr}[\mbf{\Lambda}^1]=\mu_{\mrm{A}}$ and $\mrm{Tr}[\mbf{\Lambda}^2]=\mu_{\mrm{B}}$.
The Fibonacci sequence is rewritten by $\mbf{\Lambda}^g=\mbf{\Lambda}^{g-2}\mbf{\Lambda}^{g-1}$ and thus the $g$th generation of $\mrm{Tr}[\mbf{\Lambda}^g]$ is recursively given by
\begin{equation}
\mrm{Tr}[\mbf{\Lambda}^g]=\mrm{Tr}[\mbf{\Lambda}^{g-1}]\mrm{Tr}[\mbf{\Lambda}^{g-2}]-\mrm{Tr}[\mbf{\Lambda}^{g-3}].
\label{eq:req}
\end{equation}
According to Eq.~(\ref{eq:g0g}), if $|\mrm{Tr}[\mbf{\Lambda}^g]|$ is bounded for any generation $g$, the Lyapunov exponent $\gamma_0^g$ goes to zero in the thermodynamical limit $N_g\to\infty$.
Since $\gamma_0^g=0$ means the critical condition between topological and non-topological phases, the condition corresponds to the bounded $|\mrm{Tr}[\mbf{\Lambda}^g]|$ in the thermodynamical limit.  
If we refer to the definition of the Mandelbrot set,~\cite{Mandelbrot77,note3} this non-linear recursive equation has a possiblity to produce a new type of fractal structure.
To confirm this expectation, we investigate the fractal dimension by using a box-counting approach for the intricate structure of the phase diagram.

\begin{figure}[t]
	\centering
	\includegraphics[width=0.45\textwidth]{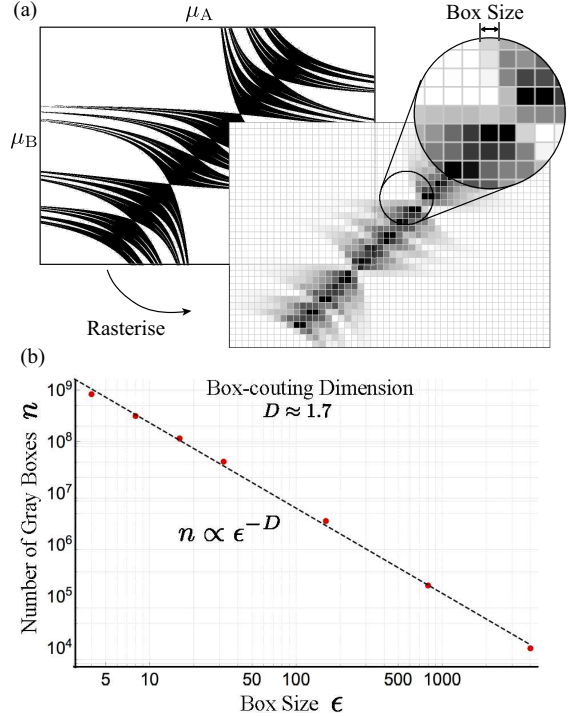}
	\caption{(Color online)(a) The region of $\gamma_0^g=0$ (black color) for the $g=17$ generation of Fibonacci chain. Rasterizing the plot with a finite box size $\epsilon$, we count the number of gray boxes $n$.
		(b) Box-counting approach. Fitting the data in the $n$-$\epsilon$ plane by $n\propto \epsilon^{-D}$ gives the box-counting dimension $D\approx 1.7$.}
	\label{fig:tpd}
\end{figure}

Figure~\ref{fig:tpd}(a) shows the critical region of the $g=17$ generation, where black (white) color represents the critical region given by $\gamma_0^g=0$ denoted as $c=1$ (the off-critical region denoted as $c=0$). 
Using a box-counting approach for Fig.~\ref{fig:tpd}(a), we calculate the box-counting dimension of the boundary of critical region as follows~\cite{Mandelbrot77}.
We firstly rasterize the figure with a coarse-graining function given by $\bar{c}_\epsilon(\bm{\mu})=\epsilon^{-2}\int_{\mcl{R}_{\epsilon}(\bm{\mu})} c(\bm{\mu}^\prime) \,\der^2\mu^\prime$, where the integrated region is defined by $\mcl{R}_\epsilon(\bm{\mu})=[\mu_{\mrm{A}}-\epsilon/2,\mu_{\mrm{A}}+\epsilon/2]\otimes[\mu_{\mrm{B}}-\epsilon/2,\mu_{\mrm{B}}+\epsilon/2]$.
If the integrated region $\mcl{R}_\epsilon(\bm{\mu})$ includes the phase boundary, the averaged criticality has an intermediate value: $0<\bar{c}_\epsilon<1$.
Therefore, we count the number of boxes $n$ that have intermediate values as a function of $\epsilon$, which is plotted in Fig.~\ref{fig:tpd}(b). 
The box-counting dimension $D\approx 1.7$ is thus found in the semi-log plot as an exponent of $n(\epsilon)\propto \epsilon^{-D}$.
Though $D$ depends on the generation $g$, we find that the extrapolated value of $D$ in the thermodynamical limit is around 1.7.
We thus conclude that the intricate structure in the critical region $\gamma_0^g=0$ exhibits a non-trivial fractal dimension.

\section{Summary and Discussion}
In conclusion, we have studied topological properties of the Fibonacci--Kitaev chain, where the Fibonacci sequence introduces quasiperiodic potential in a uniform Kitaev chain. 
In this model, we have found that the Fibonacci potential affects the topological phase diagram and makes an intricate ravine structure for the critical values of $p$-wave superconducting pairing potential separating topological and non-topological phases.
We have confirmed a self-similar fractal structure appearing in the phase boundary.
This fractal structure is expected to come from a non-linear recursive equation of the trace of transfer matrices, which determines the topological invariant.
In addition, this type of trasfer matrix was reported by Ref.~[19] with a conserved quantity in case of zero pairing potential, and thus there remains a further study on relation between the fractal structure and the conserved quantity.
Using a box-counting approach, we have determined the value of the box-counting dimension for the fractal structure to be 1.7.
We believe that the system considered here is a possible candidate to examine experimentally both fractal and topology on the same footing, by making use of, for example, recent development of ultracold atoms on quasiperiodic one-dimensional lattice~\cite{Modugno10}.
Therefore, the present study will contribute not only to bridge between a uniform system with a topology and a randomly-localized system but also to break the dawn of researches on fractal and topology.

\appendix
\section{Derivation of Equations}
\subsection{Transfer matrix}
In this subsection, we derive the transfer matrix (\ref{eq:tm}) from the original Hamiltonian (\ref{eq:ham}), provided that the Majorana zero-energy mode  commute with the Hamiltonian.
The Hamiltonian of Majorana operators $a_j=c_j^\dag+c_j$ and $b_j=\im (c_j^\dag-c_j)$ is given by,
\begin{equation}
\small
\label{eq:MajHam}
\mcl{H}=\frac{\im}{2}\sum_{j=1}^{N-1} \left[\left(1+\Delta\right)a_jb_{j+1}+\left(1-\Delta\right)a_{j+1}b_{j}\right]+\frac{\im}{2}\sum_{j=1}^{N} \mu_j a_jb_j + \mrm{const.}
\end{equation}
Here, we note that the Majorana operators obey commutation relations: $\{a_i,b_j\}=0$ and $\{a_i,a_j\}=\{b_i,b_j\}=2\delta_{ij}$.
If there is a Majorana zero-energy mode, this mode should commute with the Hamiltonian.
Assuming that the Majorana zero-energy mode is obtained by a superposition of Majorana fermions $Q_{\mrm{a}}=\sum_i\alpha_i a_i$ or $Q_{\mrm{b}}=\sum_i\beta_i b_i$ ($\alpha_i,\beta_i\in \mbb{R}$), the conditions of Majorana zero-energy mode $[\mcl{H},Q_{\mrm{a}}]=0$ and $[\mcl{H},Q_{\mrm{b}}]=0$ give following equations:
\begin{eqnarray}
\sum_{i=1}^{N-1}\left[(1+\Delta)\alpha_i b_{i+1}+(1-\Delta)\alpha_i b_{i-1}\right]+\sum_{i=1}^N\mu_i\alpha_i b_i=0,\\
\sum_{i=1}^{N-1}\left[(1+\Delta)\beta_i a_{i-1}+(1-\Delta)\beta_i a_{i}\right]+\sum_{i=1}^N\mu_i\beta_i a_i=0.
\end{eqnarray}
These equations are satisfied if the coefficients obey recursive equations,
\begin{eqnarray}
(1+\Delta)\alpha_{i-1}+(1-\Delta)\alpha_{i+1}+\mu_i\alpha_i=0,\\
(1+\Delta)\beta_{i+1}+(1-\Delta)\beta_{i-1}+\mu_i\beta_i=0,
\end{eqnarray}
for $i=0\dots N$ with a boundary condition
\begin{equation}
\alpha_0=\alpha_{N+1}=\beta_0=\beta_{N+1}=0.
\end{equation}
Therefore, we can obtain Eq.~(\ref{eq:tm}) from above equations.  

\subsection{Lyapnov exponent}
To derive Eq.\ref{eq:lyup}, we modify the transfer matrix $\mbf{A}_i$ as follows~\cite{DeGottardi13},
\begin{equation}
\tilde{\mbf{A}}_i = \sqrt{\frac{1-\Delta}{1+\Delta}} \mbf{U}^{-1} \mbf{A}_i \mbf{U} = \left(
\begin{matrix}
\frac{\mu_i}{\sqrt{1-{\it\Delta}^2}} & -1 \\ 1 & 0
\end{matrix}
\right)
\end{equation}
with
\begin{equation}
\mbf{U}=\left(\begin{matrix}
\left(\frac{1-\Delta}{1+\Delta}\right)^{\frac{1}{4}}&0\\0&	\left(\frac{1+\Delta}{1-\Delta}\right)^{\frac{1}{4}}
\end{matrix}\right).
\end{equation}
Here, the modified transfer matrix $\tilde{\mbf{A}}_i$ as a function of chemical potential $\{\mu_i\}$ and pairing potential ${\it\Delta}$ is given by the transfer matrix with a renormalized chemical potential $\left\{\frac{\mu_n}{\sqrt{1-\Delta^2}}\right\}$ and zero pairing potential,
\begin{equation}
\label{a7}
\tilde{\mbf{A}}_i(\{\mu_i\},\Delta)=\mbf{A}_i\left(\left\{\frac{\mu_n}{\sqrt{1-\Delta^2}}\right\},0\right).
\end{equation}
By the transformation, the product of transfer matrices is rewritten by 
\begin{equation}
\label{a8}
\mbf{\Lambda}^g = \prod_{i=1}^{N_g}\mbf{A}_i =\left(\frac{1-\Delta}{1+\Delta}\right)^{\frac{N_g}{2}} \mbf{U}\left(\prod_{i=1}^{N_g}\tilde{\mbf{A}}_i\right)\mbf{U}^{-1}.
\end{equation}
Since the eigenvalues of $\mbf{U}\mbf{X}\mbf{U}^{-1}$ with an arbitrary diagonalizable 2-by-2 matrix $\mbf{X}$ are the same as the eigenvalues of $\mbf{X}$, logalithm of Eq.~(\ref{a8}) gives the relationship in Eq.~(\ref{eq:lyup}).


\begin{thebibliography}{00}
\bibitem{note} Here we assume that the system given by a certain generation of Fibonacci chain is repeatedly connected and form an infinite system for explaining the topological invariant. Without any repetition of Fibonacci chain, the normalization condition is satisfied only when the eigenvalues of $\mbf{\Lambda}^g$ exponentially decrease (or increase) as increasing the system size $N_g$. This assumption of repetition is not necessary when we analyze the Lyapunov exponent $\gamma^g$, because the logarithm of the eignenvalues is divided by $N_g$ in the definition of $\gamma^g$.
\bibitem{note2} If we assume a periodicity of the Fibonacci chain, whose lattice is obtained by the cutting and projecting approach with a finite cutting depth $d_c<\infty$ in Fig.~\ref{fig:fibo}(b), there remain some possibilities for starting point of the Fibonacci chain.
We also examine the dependence of the Lyapunov exponent $\gamma^g$ on the starting point, and confirm that the maximal difference of $\gamma^g$ among the Fibonacci chains of the same generation with different starting points, is suppressed by the order $O(1/N_g)$.
\bibitem{note3} The Mandelbrot set is defined by a set of compex numbers $C_0$ which do not diverge after the non-linear recursive equation $z_{n+1}=z_n^2+C_0$ ($z_n\in\mbb{C}$ for $n\in\mbb{N}$) with $z_0=0$. However, since $\mrm{Tr}[\mbf{\Lambda}^g]$ recursively obtained in Eq.~(\ref{eq:req}) is not a complex number but a real number, the relationship between these equations is not so clear from the mathematical point of view.
\end{thebibliography}
\end{document}